\documentstyle[12pt]{article}
\begin{document}

\begin{flushright}{NUP-A-2004-3}
\end{flushright}
\vskip 0.5 truecm

\begin{center}
{\Large{\bf Path Integral for Space-time Noncommutative Field 
Theory}}
\end{center}
\vskip .5 truecm
\centerline{\bf Kazuo Fujikawa }
\vskip .4 truecm
\centerline {\it Institute of Quantum Science, College of 
Science and Technology  }
\centerline {\it Nihon University, Chiyoda-ku, Tokyo 101-8308, 
Japan}
\vskip 0.5 truecm

\makeatletter
\@addtoreset{equation}{section}
\def\theequation{\thesection.\arabic{equation}}
\makeatother

\begin{abstract}
The path integral for space-time noncommutative 
theory is formulated by means of Schwinger's action principle 
which is based on the equations 
of motion and a suitable ansatz of asymptotic conditions. The 
resulting path integral has essentially the same physical basis 
as the Yang-Feldman formulation. 
It is first shown that higher 
derivative theories are neatly dealt with by the path integral 
formulation, and the underlying canonical structure is recovered 
 by the Bjorken-Johnson-Low (BJL)  
prescription from correlation functions defined by the path 
integral. A simple theory which is non-local in time is then 
analyzed for an illustration of the complications related to 
 quantization, unitarity and positive energy conditions. 
From the view point of BJL prescription, the naive quantization 
in the interaction picture is justified for space-time 
noncommutative theory but not for the simple theory non-local  
in time.  We finally show that the perturbative 
unitarity and the positive energy condition, in the sense 
that only the positive energy flows in the positive time 
direction for any fixed time-slice in space-time, are not 
simultaneously satisfied for space-time noncommutative theory by
 the known methods of quantization. 
\end{abstract}

\section{Introduction}
The field theory in noncommutative space and time has a long 
history\cite{snyder, yang, connes, doplicher}. The quantization 
of space-time 
noncommutative theory, which contains noncommutative 
parameters in the time direction, is known to be problematic; 
the difficulties are common to those general theories non-local 
in time. The past analyses of non-local theories are found, for 
example, in \cite{hayashi, imamura, marnelius} and references
quoted  therein. First of all, no 
canonical formulation of such  theories is known since a sensible
 definition of canonical momenta is not
 known. Naturally, several authors showed the violation of
unitarity in space-time noncommutative theory\cite{gomis-mehen,
alvarez-gaume, chu}. On the other hand, it has been pointed out 
that a suitable definition of time-ordering
operation restores the unitarity in space-time noncommutative 
theory\cite{bahns, rim-yee, sibold}. In view of the conflicting 
statements in the literature, one may ask what is the sensible
definition of quantized space-time noncommutative theory, in 
particular, if the naive quantization of space-time 
noncommutative theory in the interaction picture is really 
justified. One may also  ask if the time 
ordering can be freely modified  without introducing other 
complications in space-time noncommutative theory. The main 
purpose of the present paper is to analyze these basic issues.  

To analyze the properties of quantized theory whose canonical 
quantization is not known, one needs to define a quantized 
theory in a more general setting. As one of such possibilities
it is shown that the path integral on the basis of 
 Schwinger's action principle, which is based on the formally
quantized equations of motion and a suitable ansatz of 
asymptotic conditions, provides a proper starting point of 
analyses.
The validity of this approach is similar to that of the 
Yang-Feldman formulation\cite{yang-feldman} which has been 
utilized in quantizing the noncommutative theory, but the 
time ordering operation is more rigidly specified in the path 
integral. In this path integral approach, the canonical 
structure is recovered later by means of  Bjorken-Johnson-Low
 (BJL) prescription\cite{bjorken} once one defines correlation 
functions by the path integral. 

We first illustrate that we can provide a reliable basis for the 
quantization of higher derivative theory by the path integral
described above, which may be regarded as a first step
 toward the quantization of space-time noncommutative theory. 
We show how to recover the canonical structure for higher 
derivative theory from the  path integral formulation.
We then discuss the quantization of a simple field theory 
non-local in time.
Some of the basic issues related to the quantization itself and 
the unitarity and positive energy conditions are analyzed. In 
the 
framework of BJL prescription, it is shown that the quantization
 on the basis of a naive interaction
picture is not justified if the interaction contains non-local
terms in time. The path integral quantization breaks 
perturbative unitarity, but it ensures the positive energy 
condition in the sense that only the positive energy flows in 
the positive time direction for any fixed time-slice in 
space-time by means of Feynman's $m^{2}-i\epsilon$ prescription. 
One can define a unitary S-matrix by using a  modified time 
ordering, but the positive energy condition is spoiled together 
with a smooth Wick rotation to Euclidean theory in the 
modified time ordering.

We finally analyze the quantum theory of space-time 
noncommutative theory. In this theory it is shown that the 
naive quantization in the interaction picture is justified even 
after one incorporates the higher order corrections 
perturbatively in contrast to the naive theory non-local in 
time, though this does not provide a basis for the 
non-perturbative definition of quantization. The path integral 
quantization with the 
Feynman's $m^{2}-i\epsilon$ prescription spoils the perturbative
unitarity though the positive energy condition in the sense 
that only the positive energy flows in the positive time 
direction for any fixed time-slice in space-time is ensured. 
One can define a unitary S-matrix for space-time noncommutative 
theory by using a modified time ordring but the positive energy 
condition  is spoiled together with a smooth Wick rotation to 
Euclidean theory.

\section{Higher derivative theory and canonical structure}
In this section we give a path integral formulation of 
higher derivative theory and then show how to recover the 
canonical structure from the path integral. This analysis is 
useful to understand the basis of path integrals defined by
means of Schwinger's action principle.

For simplicity, we first study the theory defined by
\begin{equation}
{\cal L}=\frac{1}{2}\partial_{\mu}\phi(x)\partial^{\mu}\phi(x)
+\lambda\frac{1}{2}\phi(x)\Box^{2}\phi(x)
\end{equation}
where\footnote{Our metric convention is 
$g_{\mu\nu}=(1,-1,-1,-1)$.}
\begin{equation}
\Box=\partial_{\mu}\partial^{\mu}
\end{equation}
and $\lambda$ is a real constant. The canonical formulation
of higher derivative theory such as the present one has been 
analyzed in \cite{nambu}, for example.

We instead start with Schwinger's action principle and 
consider the Lagrangian with a source function $J(x)$
\begin{equation}
{\cal L}_{J}=\frac{1}{2}\partial_{\mu}\phi(x)
\partial^{\mu}\phi(x)
+\lambda\frac{1}{2}\phi(x)\Box^{2}\phi(x) +J(x)\phi(x).
\end{equation}
The Schwinger's action principle starts with the equation
of motion
\begin{eqnarray}
&&\langle +\infty|-\Box\hat{\phi}(x)+\lambda\Box^{2}
\hat{\phi}(x)+J(x)
|-\infty\rangle_{J}\nonumber\\
&&=\{-\Box\frac{\delta}{i\delta J(x)}
+\lambda\Box^{2}\frac{\delta}{i\delta J(x)}+J(x)\}
\langle +\infty|-\infty\rangle_{J}=0.
\end{eqnarray}
We here assume the existence of a formally quantized field
$\hat{\phi}(x)$, though its detailed properties are not specified
yet, and the asymptotic states $|\pm\infty\rangle_{J}$
in the presence of a source function $J(x)$ localized in 
space-time. The path
integral is then defined as a formal  solution of
the above functional equation
\begin{equation}
\langle +\infty|-\infty\rangle_{J}=
\int{\cal D}\phi\exp\{i\int d^{4}x{\cal L}_{J}\}.
\end{equation}
We now define the Green's function (correlation function) by
\begin{eqnarray}
\langle +\infty|T^{\star}\hat{\phi}(x)\hat{\phi}(y)
|-\infty\rangle
&=&\frac{\delta}{i\delta J(x)}\frac{\delta}{i\delta J(y)}
\langle +\infty|-\infty\rangle_{J}|_{J=0}\nonumber\\
&=&\frac{1}{i}\frac{1}{\Box - i\epsilon - 
\lambda\Box^{2}}\delta(x-y).
\end{eqnarray}
This Green's function contains all the information about the 
quantized field.

The BJL prescription states that we can replace the covariant 
$T^{\star}$ product by the conventional $T$ product when 
\begin{equation}
\lim_{k^{0}\rightarrow\infty}\int d^{4}x e^{ik(x-y)}
\langle +\infty|T^{\star}\hat{\phi}(x)\hat{\phi}(y)
|-\infty\rangle=\lim_{k^{0}\rightarrow\infty}
\frac{i}{k^{2} + i\epsilon + \lambda(k^{2})^{2}}=0.
\end{equation}
An elementary account of the BJL prescription is given in the 
Appendix of Ref.\cite{fuji-pvn}, for example.
Thus we have
\begin{equation}
\int d^{4}x e^{ik(x-y)}
\langle +\infty|T \hat{\phi}(x)\hat{\phi}(y)
|-\infty\rangle=\frac{i}{k^{2}+ i\epsilon + \lambda(k^{2})^{2}}.
\end{equation}
By multiplying a suitable powers of the momentum variable 
$k_{\mu}$, we can recover the canonical commutation relations.
For example, 
\begin{eqnarray}
&&k_{\mu}\int d^{4}x e^{ik(x-y)}
\langle +\infty|T \hat{\phi}(x)\hat{\phi}(y)
|-\infty\rangle\nonumber\\
&&=\int d^{4}x( -i\partial^{x}_{\mu}e^{ik(x-y)})
\langle +\infty|T \hat{\phi}(x)\hat{\phi}(y)
|-\infty\rangle\nonumber\\
&&=\int d^{4}x e^{ik(x-y)}\{
\langle +\infty|T i\partial^{x}_{\mu} \hat{\phi}(x)\hat{\phi}(y)
|-\infty\rangle\nonumber\\ 
&&\ \ \ \ \ \ \ \ + i\delta(x^{0}-y^{0})
\langle +\infty|[\hat{\phi}(x),\hat{\phi}(y)]|-\infty\rangle
\}\nonumber\\
&&=\frac{ik_{\mu}}{k^{2}+ i\epsilon + \lambda(k^{2})^{2}}.
\end{eqnarray}
An analysis of this relation in the limit 
$k_{0}\rightarrow\infty$  gives
\begin{eqnarray}
&&\delta(x^{0}-y^{0})[\hat{\phi}(x),\hat{\phi}(y)]=0,\nonumber\\
&&\int d^{4}x e^{ik(x-y)}\{
\langle +\infty|T i\partial^{x}_{\mu} \hat{\phi}(x)\hat{\phi}(y)
|-\infty\rangle 
=\frac{ik_{\mu}}{k^{2}+ i\epsilon + \lambda(k^{2})^{2}}.
\end{eqnarray}
Note that the limit  $k_{0}\rightarrow\infty$ of the Fourier
transform of a $T$ product such as 
$\langle +\infty|T i\partial^{x}_{\mu} \hat{\phi}(x)\hat{\phi}(y)
|-\infty\rangle$ vanishes {\em by definition}.

By repeating the procedure with (2.10), we obtain
\begin{eqnarray}
&&\delta(x^{0}-y^{0})[\partial_{0}\hat{\phi}(x),\hat{\phi}(y)]
=0,\nonumber\\
&&\int d^{4}x e^{ik(x-y)}
\langle +\infty|T (i)^{2}\Box \hat{\phi}(x)\hat{\phi}(y)
|-\infty\rangle 
=\frac{ik^{2}}{k^{2}+ i\epsilon + \lambda(k^{2})^{2}}
\end{eqnarray}
and 
\begin{eqnarray}
&&\delta(x^{0}-y^{0})
[\partial^{2}_{0}\hat{\phi}(x),\hat{\phi}(y)]
=0,\nonumber\\
&&\int d^{4}x e^{ik(x-y)}
\langle +\infty|T (i)^{3}\partial_{\mu}\Box \hat{\phi}(x)
\hat{\phi}(y)
|-\infty\rangle 
=\frac{ik_{\mu}k^{2}}{k^{2}+ i\epsilon + \lambda(k^{2})^{2}}.
\end{eqnarray}
The final step then gives
\begin{eqnarray}
\delta(x^{0}-y^{0})
[\partial^{3}_{0}\hat{\phi}(x),\hat{\phi}(y)]
&=&\frac{i}{\lambda},\nonumber\\
\int d^{4}x e^{ik(x-y)}
\langle +\infty|T (i)^{4}\Box^{2} \hat{\phi}(x)
\hat{\phi}(y)
|-\infty\rangle 
&=&\frac{i(k^{2})^{2}}{k^{2}+ i\epsilon + \lambda(k^{2})^{2}}
-\frac{i}{\lambda}\nonumber\\
&=&-\frac{i}{\lambda}\frac{k^{2}}{k^{2}+ i\epsilon + 
\lambda(k^{2})^{2}}.
\end{eqnarray}
The last relation can be written by using (2.11) as
\begin{equation}
\int d^{4}x e^{ik(x-y)}
\langle +\infty|T \{[\lambda\Box^{2}+\Box] \hat{\phi}(x)\}
\hat{\phi}(y)|-\infty\rangle =0
\end{equation}
which is consistent with
\begin{equation}
\langle +\infty|T^{\star} 
\{[\lambda\Box^{2}+\Box] \hat{\phi}(x)\}
\hat{\phi}(y)|-\infty\rangle =i\delta^{(4)}(x-y)
\end{equation}
derived from the path integral, when combined with the 
definitions of $T$ and $T^{\star}$ products.

We thus derived the canonical commutation relations for the 
higher derivative theory
\begin{eqnarray}
&&\delta(x^{0}-y^{0})[\hat{\phi}(x),\hat{\phi}(y)]=0,\nonumber\\
&&\delta(x^{0}-y^{0})[\partial_{0}\hat{\phi}(x),\hat{\phi}(y)]
=0,\nonumber\\
&&\delta(x^{0}-y^{0})
[\partial^{2}_{0}\hat{\phi}(x),\hat{\phi}(y)]
=0,\nonumber\\
&&\delta(x^{0}-y^{0})
[\partial^{3}_{0}\hat{\phi}(x),\hat{\phi}(y)]
=\frac{i}{\lambda}\delta^{(4)}(x-y).
\end{eqnarray}
We can also confirm from (2.11) by considering the derivative 
with respect to the variable $y^{\mu}$ and by following the 
procedure similar to the above
\begin{eqnarray}
&&\delta(x^{0}-y^{0})[\partial_{0}\hat{\phi}(x),
\partial_{0}\hat{\phi}(y)]
=0,\nonumber\\
&&\delta(x^{0}-y^{0})
[\partial^{2}_{0}\hat{\phi}(x),\partial_{0}\hat{\phi}(y)]
=-\frac{i}{\lambda}\delta^{(4)}(x-y).
\end{eqnarray}
The general rule is that the commutator
\begin{equation}
[\hat{\phi}^{(m)}(x), \hat{\phi}^{(l)}(y)]\delta(x^{0}-y^{0})
\neq 0
\end{equation}
where $m+l=n-1$ for a theory with the n-th time derivative. 
Here  $\hat{\phi}^{(l)}(x)$ stands for the l-th time derivative
of $\hat{\phi}(x)$
\begin{equation}
\hat{\phi}^{(l)}(x)=\frac{\partial^{l}}{\partial(x^{0})^{l}}
\hat{\phi}(x).
\end{equation}
 We thus derive all the canonical commutation relations (2.16)
and (2.17) 
from the path integral defined by the  Schwinger's action 
principle and the $T^{\star}$ product, and those 
 commutation relations naturally agree with the relations 
derived by a canonical formulation of the higher derivative 
theory\cite{nambu}. A crucial property of the higher derivative
theory is that the canonical commutation relations are defined 
by the ``term with the highest derivative'' with the parameter
 $\lambda$. The quantization with a naive 
picture with $\lambda=0$ even for a small parameter $\lambda$ 
does not correctly describe even the qualitative features of the 
quantized theory. It is well-known that the above higher 
derivative theory contains a negative norm state, and thus not
unitary.  

We next comment on  a higher derivative theory defined by
\begin{equation}
{\cal L}_{J}=\frac{1}{2}\partial_{\mu}\phi(x)
\partial^{\mu}\phi(x)
-\lambda\phi(x)(\Box\phi(x))(\Box\phi(x)) +J(x)\phi(x).
\end{equation}
In this case, one can confirm that the one-loop diagrams (in a
naive formulation of perturbation theory)
induce a divergence corresponding to the term 
$\phi(x)(\Box^{2}\phi(x))$. This suggests that a consistent 
theory needs to be formulated at least with
\begin{equation}
{\cal L}_{J}=\frac{1}{2}\partial_{\mu}\phi(x)
\partial^{\mu}\phi(x)+\lambda_{1}\phi(x)(\Box^{2}\phi(x))
-\lambda\phi(x)(\Box\phi(x))(\Box\phi(x)) +J(x)\phi(x)
\end{equation}
with a suitable constant $\lambda_{1}$ from the beginning. We 
thus arrive at the 
case we analyzed above in (2.1). Namely, the higher derivative 
terms
in the interaction  generally lead to the problem of 
quantizing higher derivative theory. (This is also the case for
a non-renormalizable extension of the supersymmetric Wess-Zumino
model where a higher derivative K\"{a}hler term is 
induced\cite{fuji-lang}.) The canonical analysis of 
such a theory is involved, but the path integral analysis is 
relatively easier as illustrated above, in addition to giving a 
simple path integral formula for correlation 
functions defined by the $T^{\star}$ product. 

\section{Quantization of a  theory non-local in time}

We examine a non-local theory defined by 
\begin{eqnarray}
{\cal L}_{J}&=&-\frac{1}{2}\phi(x)\Box
[\phi(x+\xi)+\phi(x-\xi)] + J(x)\phi(x)
\nonumber\\
&=&-\frac{1}{2}\phi(x)\Box 
[e^{i\xi\hat{p}}+e^{-i\xi\hat{p}}]\phi(x) + J(x)\phi(x).
\end{eqnarray}
This Lagrangian is somewhat analogous to a lattice theory, but 
we treat this Lagrangian as a non-local theory defined in 
continuum. A formal integration of the Schwinger's action 
principle
\begin{eqnarray}
&&\langle +\infty|-\Box[\hat{\phi}(x+\xi)+\hat{\phi}(x-\xi)] 
+J(x)|-\infty\rangle_{J}\nonumber\\
&&=\{-\Box[\frac{\delta}{i\delta J(x+\xi)}
+\frac{\delta}{i\delta J(x-\xi)}]+J(x)\}
\langle +\infty|-\infty\rangle_{J}=0
\end{eqnarray}
gives a path integral
\begin{eqnarray}
\langle +\infty|-\infty\rangle_{J}=
\int{\cal D}\phi\exp\{i\int d^{4}x{\cal L}_{J}\},
\end{eqnarray}
which in turn leads to the correlation function
\begin{equation}
\langle T^{\star}\hat{\phi}(x)\hat{\phi}(y)\rangle =
\int\frac{d^{4}k}{(2\pi)^{4}}\frac{i}{(k^{2}+ i\epsilon)
[e^{ik\xi}+e^{-ik\xi}
]}
e^{-ik(x-y)}
\end{equation}
or
\begin{equation}
\int d^{4}x e^{ik(x-y)}
\langle T^{\star}\hat{\phi}(x)\hat{\phi}(y)\rangle =
\frac{i}{(k^{2}+ i\epsilon)[e^{ik\xi}+e^{-ik\xi}]}.
\end{equation}
For a time-like vector $\xi$, which may be chosen as 
$(\xi^{0},0,0,0)$,  the right-hand side of this
expression multiplied by any power of $k_{0}$ goes to zero 
\begin{equation}
\lim_{k_{0}\rightarrow i\infty}
\frac{i(k_{0})^{n}}{(k^{2}+ i\epsilon)[e^{ik\xi}+e^{-ik\xi}]}=0
\end{equation}
for $k_{0}$ along the imaginary axis 
in the complex $k_{0}$ plane\footnote{For general cases, we take
 $k_{0}$ along the imaginary axis  as is suggested by a smooth 
Wick rotation to avoid  possible singularities; for a 
theory analyzed in the previous section, this  careful choice of
 the direction of $k_{0}$ was not required. }. Thus the 
application of BJL
prescription leads to (for any pair of non-negative integers $n$ 
and $m$) 
\begin{equation}
[\hat{\phi}^{(n)}(x), \hat{\phi}^{(m)}(y)]\delta(x^{0}-y^{0})=0
\end{equation}
where $\hat{\phi}^{(n)}(x)$ stands for the n-th time derivative
of $\hat{\phi}(x)$ as in (2.19).
 This relation is consistent with the $N\rightarrow\infty$ 
limit of a  higher derivative theory obtained by a truncation 
of the power series exapansion of $e^{\pm i\xi\hat{p}}$ at the 
$N$-th power in the starting Lagrangian (3.1). See also the 
analysis in the previous section.  
In contrast, for a space-like $\xi$ for which one may choose 
$\xi=(0,\vec{\xi})$, one recovers the result of the naive 
canonical quantization of (3.1) 
\begin{eqnarray}
&&\delta(x^{0}-y^{0})[\hat{\phi}(x),\hat{\phi}(y)]=0,\nonumber\\
&&\delta(x^{0}-y^{0})[\partial_{0}\hat{\phi}(x),\hat{\phi}(y)]
=\frac{-i}{[e^{i\xi\hat{p}}+e^{-i\xi\hat{p}}]}\delta^{(4)}(x-y)
\end{eqnarray}
by means of BLJ prescription; in the right-hand side of (3.8), 
$\hat{p}$ stands for the (spatial) momentum operator acting on 
the coordinate $\vec{x}$. 

We can thus define no 
sensible canonical structure for the present non-local theory
for a time-like $\xi$.
Nevertheless, we can fomally define a quantum theory by the
Schwinger's action principle and the path integral.
The quantization is defined by a specification of 
$\langle T^{\star}\hat{\phi}(x)\hat{\phi}(y)\rangle$.  

We next analyze a theory which contains a non-local interaction 
\begin{eqnarray}
{\cal L}&=&\frac{1}{2}\partial_{\mu}\phi(x)\partial^{\mu}\phi(x)
-\frac{m^{2}}{2}\phi(x)\phi(x)\nonumber\\
&& 
-\frac{g}{2}[\phi(x+\xi)\phi(x)\phi(x-\xi)+
\phi(x-\xi)\phi(x)\phi(x+\xi)]
+\phi(x) J(x)\nonumber\\
&=&\frac{1}{2}\partial_{\mu}\phi(x)\partial^{\mu}\phi(x)
-\frac{m^{2}}{2}\phi(x)\phi(x)\nonumber\\
&& 
-\frac{g}{2}[(e^{\xi^{\mu}\partial_{\mu}}\phi(x))
\phi(x)(e^{-\xi^{\mu}\partial_{\mu}}\phi(x))
+(e^{-\xi^{\mu}\partial_{\mu}}\phi(x))\phi(x)
(e^{\xi^{\mu}\partial_{\mu}}\phi(x))]\nonumber\\
&&+\phi(x) J(x)
\end{eqnarray}
where $\xi^{\mu}$ is a constant four-vector. This theory is 
not Lorentz invariant because of the constant vector $\xi^{\mu}$.
When one chooses $\xi^{\mu}$ to be a time-like vector
$\xi^{2}=(\xi^{0})^{2}-(\vec{\xi})^{2}>0$,
the quantization of the above theory  is analogous to that of 
 space-time noncommutative theory. When one works in the frame
\begin{equation}
\xi^{\mu}=(\xi^{0},0,0,0)\ \  {\rm with}\ \ \xi^{0}>0
\end{equation}
which we adopt in the rest of this section, it is obvious that 
the unitary time development (in the sense
of the Schr\"{o}dinger equation) for the small time interval 
$\Delta t<\xi^{0}$ is not defined. One may examine a naive 
Hamiltonian
\begin{eqnarray}
{\cal H}&=&\frac{1}{2}\Pi^{2}(x)
+\frac{1}{2}\vec{\nabla}\phi(x)\vec{\nabla}\phi(x)    
+\frac{1}{2}m^{2}\phi^{2}(x)\nonumber\\
&& +\frac{g}{2}[\phi(x-\xi)\phi(x)\phi(x+\xi)+
\phi(x+\xi)\phi(x)\phi(x-\xi)]
\end{eqnarray}
where $x^{\mu}=(0,\vec{x})$ and $\xi^{\mu}=(\xi^{0},\vec{0})$
, and $\Pi(x)=\frac{\partial}{\partial x^{0}}\phi(x)$ is a naive
 canonical momentum conjugate to $\phi(x)$.
This Hamiltonian is formally hermitian, 
${\cal H}^{\dagger}={\cal H}$, but ${\cal H}$ is not local in 
the time coordinate and does not generate time development in 
the conventional sense for 
the small time interval $\Delta t<\xi^{0}$. The equal-time 
commutation relation, for example,
\begin{eqnarray}
[\int d^{3}x {\cal H}(x), \phi(y)]\delta(x^{0}-y^{0})
\end{eqnarray}
is not well defined, since 
$[\phi(x+\xi), \phi(y)]\delta(x^{0}-y^{0})$ is not well specified
in the non-perturbative level. 

Nevertheless, one may study the path integral quantization 
without specifying  the precise quantization condition of field 
variables. This aspect is analogous to the Yang-Feldman 
formulation.
One may thus  define a path integral by means of  Schwinger's 
action  principle and a suitable ansatz of asymptotic conditions
as in (3.2)
\begin{eqnarray}      
\langle +\infty|-\infty\rangle_{J}=\int{\cal D}\phi
\exp[i\int d^{4}x{\cal L}_{J}].
\end{eqnarray}
One may then define a formal expansion in powers of the coupling
constant $g$. It is interesting to examine what one learns as
to the canonical quantization and unitarity relations defined by
 the Feynman diagrams.

We study one-loop diagrams in a formal perturbative expansion 
in powers of the coupling constant $g$ by starting with a 
tentative ansatz of quantization
\begin{equation}
\langle T^{\star}\hat{\phi}(x)\hat{\phi}(y)\rangle=
\frac{-i}{\Box + m^{2}-i\epsilon}\delta(x-y)
=\int\frac{d^{4}k}{(2\pi)^{4}}
e^{-ik(x-y)}\frac{i}{k^{2}-m^{2}+i\epsilon} 
\end{equation}
which is equivalent to a canonical quantization of free theory.
One-loop self-energy diagrams contain the contributions
\begin{eqnarray}
&&\frac{(-ig)^{2}}{2}
\int d^{4}x d^{4}y \phi(x)\langle T^{\star}\phi(x+\xi)
\phi(x-\xi)\phi(y+\xi)\phi(y-\xi)\rangle \phi(y)\nonumber\\
&&=\frac{(-ig)^{2}}{2}\int d^{4}x d^{4}y \phi(x)\phi(y)
[\langle T^{\star}\phi(x+\xi)\phi(y+\xi)\rangle\langle T^{\star}
\phi(x-\xi)\phi(y-\xi)\rangle\nonumber\\
&& \ \ \ \ +\langle T^{\star}\phi(x+\xi)\phi(y-\xi)\rangle
\langle T^{\star}\phi(x-\xi)\phi(y+\xi)\rangle]
\end{eqnarray}
The first term in (3.15) gives rise to a logarithmically 
divergent local 
contribution, which is absorbed into the mass renormalization, 
and the second term gives a finite non-local (approximately 
separated by $\sim 2\xi$) term. We also have  contributions
\begin{eqnarray}
&&\frac{(-ig)^{2}}{2}
\int d^{4}x d^{4}y \phi(x)\langle T^{\star}\phi(x+\xi)
\phi(x-\xi)\phi(y)\phi(y-\xi)\rangle \phi(y+\xi)\nonumber\\
&&=\frac{(-ig)^{2}}{2}\int d^{4}x d^{4}y \phi(x)\phi(y)
[\langle T^{\star}\phi(x+\xi)\phi(y-\xi)\rangle\langle
T^{\star}\phi(x-\xi)\phi(y-2\xi)\rangle\nonumber\\
&& \ \ \ \ +\langle T^{\star}\phi(x+\xi)\phi(y-2\xi)\rangle
\langle T^{\star}\phi(x-\xi)\phi(y-\xi)\rangle]
\end{eqnarray}
which contains the finite non-local terms separated  
up to the order of $\sim 3\xi$.

The first term in (3.16), for example,   gives rise to 
\begin{eqnarray}
&&g^{2}i\Sigma(k, \xi)\nonumber\\
&=&\frac{- g^{2}}{2}
\int\frac{d^{4}k}{(2\pi)^{4}}[e^{2ik\xi}e^{i\xi(p-k)} 
+ e^{-2ik\xi}e^{-i\xi(p-k)}]\frac{i}{k^{2}-m^{2}+i\epsilon}
\frac{i}{(p-k)^{2}-m^{2}+i\epsilon}
\nonumber\\
&=&\frac{- g^{2}}{2}\int\frac{d^{4}k}{(2\pi)^{4}}[e^{i\xi(p+k)} 
+ e^{-i\xi(p+k)}]\frac{i}{k^{2}-m^{2}+i\epsilon}
\frac{i}{(p-k)^{2}-m^{2}+i\epsilon}
\nonumber\\
&=&\frac{- g^{2}}{2}\int\frac{d^{4}k}{(2\pi)^{4}}[e^{i\xi(p+k)} 
+ e^{-i\xi(p+k)}]\int_{0}^{\infty}dz_{1}dz_{2}
e^{i z_{1}[k^{2}-m^{2}+i\epsilon]+iz_{2}
[(p-k)^{2}-m^{2}+i\epsilon]}.
\end{eqnarray}
Note that the Feynman's $m^{2}-i\epsilon$ prescrition 
provides a convergent factor at $z_{1,2}=\infty$.
One can further evaluate this by setting $z_{1}=\alpha x$ and 
$z_{2}=\alpha(1-x)$ as 
\begin{eqnarray}
&&g^{2}i\Sigma(k, \xi)\nonumber\\
&=&\frac{- g^{2}}{2}\int\frac{d^{4}k}{(2\pi)^{4}}
\int_{0}^{\infty}\alpha
d\alpha\int_{0}^{1}dx
e^{i\alpha[k^{2}+x(1-x)p^{2}-m^{2}+i\epsilon)]}\nonumber\\
&&\times [e^{i(2-x)\xi p+i\xi k} + e^{-i(2-x)\xi p-i\xi k}]
\nonumber\\
&=&\frac{ig^{2}}{2(4\pi)^{2}}
\int_{0}^{1}dx[e^{i(2-x)\xi p} + e^{-i(2-x)\xi p}]
\int_{0}^{\infty}
\frac{d\alpha}{\alpha}e^{i\alpha[x(1-x)p^{2}-m^{2}+i\epsilon)]
-i\frac{\xi^{2}}{4\alpha}}.
\end{eqnarray}
We analyze the part of the above amplitude
\begin{eqnarray}
\int_{0}^{\infty}
\frac{d\alpha}{\alpha}e^{i\alpha[x(1-x)p^{2}-m^{2}+i\epsilon)]
-i\frac{\xi^{2}}{4\alpha}}.
\end{eqnarray}
Following the conventional approach, we define the integral
for a Euclidean momentum $p_{\mu}$, for which $p^{2}<0$.
In this case, one can deform the integration contour along the 
negative real axis as
\begin{eqnarray}
&&\int_{0}^{\infty e^{-i\pi}}
\frac{d\alpha}{\alpha}e^{i\alpha[x(1-x)p^{2}-m^{2}+i\epsilon)]
-i\frac{\xi^{2}}{4\alpha}}\nonumber\\
&&=\int_{0}^{\infty e^{-i\pi}}
\frac{d\alpha}{\alpha}e^{-i\frac{
\sqrt{(-\xi^{2})[-x(1-x)p^{2}+m^{2}-i\epsilon)}}{2}[\alpha-
\frac{1}{\alpha}]}\nonumber\\
&&=-i\pi H^{(2)}_{0}(-i
\sqrt{(-\xi^{2})[-x(1-x)p^{2}+m^{2}-i\epsilon)})
\end{eqnarray}
for $\xi^{2}<0$. Here $H^{(2)}_{0}(z)$ stands for the Hankel
function which has an asymptotic expansion for $|z|\rightarrow
\infty$
\begin{eqnarray}
H^{(2)}_{0}(z)\sim \sqrt{\frac{2}{\pi z}}e^{-i(z-\frac{\pi}{4})}
\end{eqnarray}
for $-2\pi <{\rm arg} z <\pi$.

We thus find that for $p_{0}\rightarrow i\infty$ 
\begin{eqnarray}
\int_{0}^{\infty}
\frac{d\alpha}{\alpha}e^{i\alpha[x(1-x)p^{2}-m^{2}+i\epsilon)]
-i\frac{\xi^{2}}{4\alpha}}\sim
-\pi \sqrt{\frac{2}{\pi z}}e^{-z}
\end{eqnarray}
with
\begin{equation}
z=\sqrt{(-\xi^{2})[-x(1-x)p^{2}+m^{2}-i\epsilon)}
\end{equation}
for a space-like $\xi$, $\xi^{2}<0$. On the other hand, we have 
a damping oscillatory behavior for 
$p_{0}\rightarrow i\infty$, 
\begin{eqnarray}
\int_{0}^{\infty}
\frac{d\alpha}{\alpha}e^{i\alpha[x(1-x)p^{2}-m^{2}+i\epsilon)]
-i\frac{\xi^{2}}{4\alpha}}\sim
-i\pi \sqrt{\frac{2}{\pi z}}e^{(- iz+i\frac{\pi}{4})}
\end{eqnarray}
with
\begin{equation}
z=\sqrt{(\xi^{2})[-x(1-x)p^{2}+m^{2}-i\epsilon)}
\end{equation}
for a time-like $\xi$, $\xi^{2}>0$, which is defined by an 
analytic continuation.

When one writes the (complete) connected two-point correlation 
function with one-loop corrections as 
\begin{equation}
\langle T^{\star}\hat{\phi}(x)\hat{\phi}(y)\rangle_{ren}
=\int\frac{d^{4}p}{(2\pi)^{4}}
e^{-ip(x-y)}\frac{i}{p^{2}+g^{2}\Sigma(p, \xi)-m_{r}^{2}
+i\epsilon}, 
\end{equation}
the two-point function generally contains the
 non-local term in $g^{2}\Sigma(p, \xi)$. 
When one applies the BJL prescription to the two-point 
correlation function in a conventional local renormalizable 
theory, the 
higher order corrections do  not modify the canonical structure 
since we apply the BJL prescription to the two-point function 
 with the ultraviolet 
cut-off of  loop momenta kept fixed. In the present context
this corresponds to the replacement 
\begin{eqnarray}
&&\int_{0}^{\infty}
\frac{d\alpha}{\alpha}e^{i\alpha[x(1-x)p^{2}-m^{2}+i\epsilon)]
-i\frac{\xi^{2}}{4\alpha}}\nonumber\\
&&\rightarrow 
\int_{0}^{\infty}
\frac{d\alpha}{\alpha}e^{i\alpha[x(1-x)p^{2}-m^{2}+i\epsilon)]
-i\frac{\xi^{2}}{4\alpha}}
-\int_{0}^{\infty}
\frac{d\alpha}{\alpha}e^{i\alpha[x(1-x)p^{2}-M^{2}+i\epsilon)]
-i\frac{\xi^{2}}{4\alpha}}
\end{eqnarray}
where $M$ stands for the Pauli-Villars-type cut-off mass.

In the present case also, $g^{2}\Sigma(p, \xi)$ assumes  real
values and $g^{2}\Sigma(p, \xi)\rightarrow 0$
for $p_{0}\rightarrow i\infty$ for the space-like $\xi$ as in 
(3.22), for 
which we may take $\xi=(0, \vec{\xi})$. The canonical structure 
is not modified by the one-loop effects of  the  interaction 
non-local in the spatial distance.   

In contrast, for the time-like $\xi$ for which we may take 
$\xi=(\xi^{0}, \vec{0})$, $g^{2}\Sigma(p, \xi)$ diverges 
exponentially for $p_{0}\rightarrow i\infty$. This arises from
the behavior of the factor   
\begin{eqnarray}
[e^{i(2-x)\xi p} + e^{-i(2-x)\xi p}]
\end{eqnarray}
in (3.18) for $p_{0}\rightarrow i\infty$ and 
$\xi=(\xi^{0}, \vec{0})$, which dominates the damping 
oscillatory behavior (3.24)\footnote{The non-vanishing imaginary
part of $g^{2}\Sigma(p, \xi)$ in (3.24) for the Euclidean 
momentum given by $p_{0}\rightarrow i\infty$ is associated with 
the violation of unitarity in the present theory non-local in 
time~\cite{gomis-mehen, alvarez-gaume, chu}.}. The canonical 
structure specified 
by the BJL analysis is thus completely modified by the one-loop 
effects of  the  interaction non-local in time. After one-loop 
corrections, we essentially have the same result (3.6) as for 
the non-local theory (3.1). 
The naive ansatz of the two-point correlation function at the 
starting point of  perturbation theory (3.14) is not justified. 
We thus conclude that the present model for a time-like $\xi$
does not accommodate a consistent canonical structure of 
quantized theory. In contrast, the naive ansatz (3.14) is not 
modified by the one-loop quantum corrections for a space-like 
$\xi$.

Nevertheless, it is instructive to examine the formal 
perturbative unitarity of an S-matrix defined for the theory 
non-local in time. One may first observe that 
\begin{eqnarray}
S(t_{+}, t_{-})=e^{i\hat{H}_{0}t_{+}}e^{-i\hat{H}(t_{+}-t_{-})}
e^{-i\hat{H}_{0}t_{-}}
\end{eqnarray}
for ${\cal H}$ in (3.11) with
\begin{equation}
H_{0}=\int d^{3}x[\frac{1}{2}\Pi^{2}(0,\vec{x})
+\frac{1}{2}\vec{\nabla}\phi(0,\vec{x})\vec{\nabla}
\phi(0,\vec{x})    
+\frac{1}{2}m^{2}\phi^{2}(0,\vec{x})]
\end{equation}
is unitary
\begin{equation}
S(t_{+}, t_{-})^{\dagger}S(t_{+}, t_{-})
=S(t_{+}, t_{-})S(t_{+}, t_{-})^{\dagger}=1
\end{equation}
The formal power series expansion in the coupling 
constant\footnote{We use the notation $T_{\star}$ for the time 
ordering in non-local theory, whereas $T$ or $T^{\star}$ is 
used for the conventional time ordering with respect to 
the time variable of {\em each} field variable $\phi(x)$. }
\begin{eqnarray}
S(t_{+}, t_{-})=\sum_{n=0}^{\infty}\frac{(-i)^{n}}{n!}
\int_{t_{-}}^{t_{+}}dt_{1}....\int_{t_{-}}^{t_{+}}dt_{n}
T_{\star}(\hat{H}_{I}(t_{1})....\hat{H}_{I}(t_{n}))
\end{eqnarray}
with a hermitian 
\begin{equation}
\hat{H}_{I}(t)=e^{i\hat{H}_{0}t}\int d^{3}x
\frac{g}{2}[\phi(-\xi,\vec{x})\phi(0,\vec{x})\phi(\xi,\vec{x})
+\phi(\xi,\vec{x})\phi(0,\vec{x})\phi(-\xi,\vec{x})]
e^{-i\hat{H}_{0}t}
\end{equation}
thus defines a unitary operator
\begin{equation}
\hat{S}=\lim_{t_{-}\rightarrow -\infty, t_{+}\rightarrow +\infty}
S(t_{+}, t_{-}).
\end{equation}
This definition of a unitary operator corresponds to the 
definition of a unitary S-matrix for space-time noncommutative 
theory proposed in \cite{bahns, rim-yee}. 

It is important to recognize that the time-ordering in the 
present context is defined with repect to the time variable
appearing in $\hat{H}_{I}(t)$; if one performs a time-ordering
with repect to the time variable appearing in each field variable
$\phi(x)$, one generally obtains different results due to the 
non-local structure of the interaction term in time. Since the 
operator
$\hat{S}$ defined above is manifestly unitary, the non-unitary 
result in the conventional Feynman rules, which are based on the
 time-ordering of each operator $\phi(x)$, arises from this 
difference of time ordering. In any case, it should be possible 
to understand the origin of unitary or non-unitary S-matrix in 
the coordinate representation without recourse to the momentum 
representation of Feynman diagrams. 

When one defines 
\begin{eqnarray}
A_{1}&=&\int_{-\infty}^{\infty} dt \hat{H}_{I}(t),\nonumber\\ 
A_{2}&=&\frac{1}{2}\int_{-\infty}^{\infty} 
\int_{-\infty}^{\infty} 
dt_{1}dt_{2}T_{\star}\hat{H}_{I}(t_{1})\hat{H}_{I}(t_{2})
\nonumber\\
&=&\int_{-\infty}^{\infty} 
\int_{-\infty}^{\infty} 
dt_{1}dt_{2}\theta(t_{1}-t_{2})
\hat{H}_{I}(t_{1})\hat{H}_{I}(t_{2})
\end{eqnarray}
the unitarity relation of the above S-matrix in the second order
 of the coupling constant requires( see, for example, 
\cite{rim-yee})
\begin{eqnarray}
A_{2}+A_{2}^{\dagger}=A_{1}^{\dagger}A_{1}=A_{1}^{2}.
\end{eqnarray}
To be explicit
\begin{eqnarray}
A_{2}+A_{2}^{\dagger}&=&\int_{-\infty}^{\infty} 
\int_{-\infty}^{\infty} 
dt_{1}dt_{2}\theta(t_{1}-t_{2})\{
\hat{H}_{I}(t_{1})\hat{H}_{I}(t_{2})
+\hat{H}_{I}(t_{2})\hat{H}_{I}(t_{1})\}\nonumber\\
&=&\int_{-\infty}^{\infty} 
\int_{-\infty}^{\infty} 
dt_{1}dt_{2}\{\theta(t_{1}-t_{2})
\hat{H}_{I}(t_{1})\hat{H}_{I}(t_{2})
+\theta(t_{2}-t_{1})\hat{H}_{I}(t_{1})\hat{H}_{I}(t_{2})\}
\nonumber\\
&=&\int_{-\infty}^{\infty} \int_{-\infty}^{\infty} 
dt_{1}dt_{2}\hat{H}_{I}(t_{1})\hat{H}_{I}(t_{2})\nonumber\\
&=&\int_{-\infty}^{\infty}  dt_{1}\hat{H}_{I}(t_{1})
\int_{-\infty}^{\infty}dt_{2}\hat{H}_{I}(t_{2})\nonumber\\
&=&A_{1}^{2}
\end{eqnarray}
by noting $\theta(t_{1}-t_{2})+\theta(t_{2}-t_{1})=1$, as 
required by the unitarity relation.

In contrast, if one uses the conventional time ordering one has
\begin{eqnarray} 
A_{2}&=&\frac{1}{2}\int_{-\infty}^{\infty} 
\int_{-\infty}^{\infty} 
dt_{1}dt_{2}T^{\star}\hat{H}_{I}(t_{1})\hat{H}_{I}(t_{2})
\nonumber\\
&\neq&\int_{-\infty}^{\infty} 
\int_{-\infty}^{\infty} 
dt_{1}dt_{2}\theta(t_{1}-t_{2})
\hat{H}_{I}(t_{1})\hat{H}_{I}(t_{2})
\end{eqnarray}
since the time ordering by $T^{\star}$ is defined with respect 
to the time variable of each field $\phi(t,\vec{x})$, and thus 
the unitarity of the conventional operator 
\begin{eqnarray}
\hat{S}=\sum_{n=0}^{\infty}\frac{(-i)^{n}}{n!}
\int_{-\infty}^{+\infty}dt_{1}....\int_{-\infty}^{+\infty}
dt_{n}T^{\star}(\hat{H}_{I}(t_{1})....\hat{H}_{I}(t_{n}))
\end{eqnarray}
is not satisfied for the non-local $\hat{H}_{I}(t)$ in general.
Note that the perturbative expansion with the $T^{\star}$
product is directly defined by the path integral without 
recourse to the expression such as (3.29).

On the other hand, the positive energy condition,
which is ensured by the Feynman propagator, is  not obvious for 
the propagator defined by $T_{\star}$.
 To be specific, we have the following correlation function in 
the Wick-type reduction of the S-matrix
\begin{eqnarray}
&&\langle0|T\phi(x-\xi)\phi(y+\xi)|0\rangle\nonumber\\
&&=
\int\frac{d^{4}k}{(2\pi)^{4}}\exp[-ik((x-\xi)-(y+\xi))]
\frac{i}{k_{\mu}k^{\mu}-m^{2}+i\epsilon}\nonumber\\
&&=\theta((x-\xi)^{0}-(y+\xi)^{0})
\int\frac{d^{3}k}{(2\pi)^{3}2\omega}
\exp[-i\omega((x-\xi)^{0}-(y+\xi)^{0})+
i\vec{k}(\vec{x}-\vec{y})]\nonumber\\
&&+\theta((y+\xi)^{0}-(x-\xi)^{0})
\int\frac{d^{3}k}{(2\pi)^{3}2\omega}
\exp[-i\omega((y+\xi)^{0}-(x-\xi)^{0})+
i\vec{k}(\vec{y}-\vec{x})]
\nonumber\\
\end{eqnarray}
with $\omega=\sqrt{\vec{k}^{2}+m^{2}}$ for the conventional 
Feynman prescription with $m^{2}-i\epsilon$, which ensures that 
the positive frequency components propagate in the forward time 
direction and the negative frequency components propagate in 
the backward time direction and thus the positive energy  
flows always in the forward time 
direction. The Wick rotation to Euclidean theory in the momentum
 space is also smooth
in this prescription. The path integral with respect to the 
field variable $\phi(x)$ gives this time ordering
 or the $T^{\star}$ product.

In comparison, the non-local prescription (3.32) gives the 
following correlation function for the quantized free field in 
the Wick-type reduction
\begin{eqnarray}
&&\langle0|T_{\star}\phi(x-\xi)\phi(y+\xi)|0\rangle\nonumber\\
&&=\theta(x^{0}-y^{0})
\int\frac{d^{3}k}{(2\pi)^{3}2\omega}
\exp[-i\omega((x-\xi)^{0}-(y+\xi)^{0}))+
i\vec{k}(\vec{x}-\vec{y})]\nonumber\\
&&+\theta(y^{0}-x^{0})
\int\frac{d^{3}k}{(2\pi)^{3}2\omega}
\exp[-i\omega((y+\xi)^{0}-(x-\xi)^{0})+
i\vec{k}(\vec{y}-\vec{x})]
\nonumber\\
\end{eqnarray}
where the time-ordering step function $\theta(x^{0}-y^{0})$,
for example, and the signature of the time variable 
$(x-\xi)^{0}-(y+\xi)^{0}$ appearing in the exponential are
not correlated, and it 
 allows the negative energy to propagate in the forward
time direction also. This result is not reproduced by the 
Feynman's $m^{2} - i\epsilon$ prescription. When one considers 
an arbitrary fixed time-slice in 4-dimesional space-time, the 
condition that
all the particles crossing the time-slice carry the positive 
energy in the forward
time direction, which is regarded as the positive energy 
condition in the path integral formulation\cite{spin-sta} 
(or in perturbation 
theory in general), is not satisfied~\footnote{The present 
theory 
is a generalization of the globally unstable $\phi^{3}$ 
potential. The positive energy condition we are discussing is 
independent of this global structure of the potential, and our 
analysis is valid for the $\phi^{4}$-type potential also.}. 
This 
positive energy condition  is crucial in the analysis of 
spin-statistics theorem\cite{spin-sta, pauli}, for example.
See also \cite{chaichian} for an analysis of spin-statistics 
theorem in noncommutative theory.

We thus summarize the analysis of  this section as follows:
The naive canonical quantization in a perturbative sense is not 
justified in the present theory non-local in time
when one incorporates the higher order corrections. The unitarity
of the (formal) perturbative S-matrix is ensured if one adopts 
the $T_{\star}$ product, but the positive energy condition is not
satisfied by this prescription. Also the Wick rotation is not
obvious in this modified $T_{\star}$ product. On the other hand,
the unitarity of the $S$ matrix is spoiled if one adopts the 
conventional $T$ or $T^{\star}$ product which is defined by the 
path integral, though the positive energy condition and a 
smooth Wick rotation are ensured. 
 
\section{Quantization of space-time noncommutative theory}
We study the simplest noncommutative theory defined by
\begin{eqnarray}
{\cal L}_{J}
&=&\frac{1}{2}\partial_{\mu}\phi(x)\star\partial^{\mu}\phi(x)
-\frac{m^{2}}{2}\phi(x)\star\phi(x)\nonumber\\
&& 
-\frac{g}{3!}\phi(x)\star\phi(x)\star\phi(x)
+\phi(x)\star J(x)
\end{eqnarray}
where the $\star$ product is defined by the so-called Moyal
product
\begin{equation}
\phi(x)\star\phi(x)
=e^{\frac{i}{2}\xi\partial^{x}_{\mu}
\theta^{\mu\nu}\partial^{y}_{\nu}}
\phi(x)\phi(y)|_{y=x}
=e^{\frac{i}{2}\xi\partial_{x}\wedge\partial_{y}}
\phi(x)\phi(y)|_{y=x}.
\end{equation}
The real positive parameter $\xi$ stands for the deformation 
parameter,
and the antisymmetric parameter 
$\theta^{\mu\nu}=-\theta^{\nu\mu}$ corresponds to  
$i\xi\theta^{\mu\nu}=[\hat{x}^{\mu},\hat{x}^{\nu}]$ ; since this
 theory is  not Lorentz covariant we consider the case 
$\theta^{0i}=-\theta^{i0}\neq 0$ for a suitable $i$
but all others $\theta^{\mu\nu}=0$ in the following.

The formal quantum equation of motion is given by 
\begin{equation}
-\Box\hat{\phi}(x)-m^{2}\hat{\phi}(x)
-\frac{g}{2!}\hat{\phi}(x)\star\hat{\phi}(x)
+J(x)=0.
\end{equation}
The Yang-Feldman formulation solves this operator equation (with
$J=0$) by
imposing  suitable boundary conditions at $t=\pm\infty$ and by 
using the corresponding  two-point (free) Green's functions. It 
may be
noted that the validity of these boundary conditions is  not 
obvious in the present noncommutative theory with 
$\theta^{0i}\neq 0$.

The Schwinger's action principle starts with the relation
\begin{eqnarray}  
&&\langle +\infty|-\Box\hat{\phi}(x)-m^{2}\hat{\phi}(x)
-\frac{g}{2!}\hat{\phi}(x)\star\hat{\phi}(x)
+J(x)|-\infty\rangle_{J} \nonumber\\
&&=[-\Box\frac{\delta}{i\delta J(x)}
-m^{2}\frac{\delta}{i\delta J(x)}
-\frac{g}{2!}\frac{\delta}{i\delta J(x)}\star
\frac{\delta}{i\delta J(x)}
+J(x)]\nonumber\\
&&\ \ \times\langle +\infty|-\infty\rangle_{J}=0.
\end{eqnarray}
This relation assumes the existence of the asymptotic 
states $|\pm\infty\rangle_{J}$ at $t=\pm\infty$ in the presence 
of the source function
$J(x)$ which has a support in the finite space-time domain. This
 Schwinger's action 
principle thus depends on essentially the same set of assumptions
as those of the Yang-Feldman formulation.

The path integral is then defined as a formal integral of the 
Schwinger's action principle (4.4)
\begin{eqnarray}      
\langle +\infty|-\infty\rangle_{J}=\int{\cal D}\phi
\exp[i\int d^{4}x{\cal L}_{J}]
\end{eqnarray}
with a ``translational invariant'' path integral measure
\begin{equation}
{\cal D}(\phi+\epsilon)= {\cal D}\phi
\end{equation}
where $\epsilon(x)$ is an arbitrary infinitesimal function
independent of $\phi(x)$.
The condition (4.6) ensures that the Feynman path integral 
formula satisfies the Schwinger's action principle.  

It has been argued \cite{chepelev} that the present theory is 
renormalizable in the formal perturbative expansion in powers of
 the coupling constant $g$ starting with 
\begin{equation}
\langle T^{\star}\hat{\phi}(x)\hat{\phi}(y)\rangle=
\frac{-i}{\Box + m^{2}-i\epsilon}\delta(x-y)
=\int\frac{d^{4}k}{(2\pi)^{4}}
e^{-ik(x-y)}\frac{i}{k^{2}-m^{2}+i\epsilon}
\end{equation}
which is equivalent to a canonical quantization of free theory.
The one-loop self-energy is given by
\begin{eqnarray}
&&g^{2}i\Sigma(p, \xi)\nonumber\\
&&=\frac{-g^{2}}{2}\int \frac{d^{4}k}{(2\pi)^{4}} 
\cos^{2}(\frac{\xi}{2}p\wedge k)\frac{i}{
((p-k)^{2}-m^{2}+i\epsilon)}\frac{i}{(k^{2}-m^{2}+i\epsilon)}
\nonumber\\
&&=\frac{g^{2}}{4}\int \frac{d^{4}k}{(2\pi)^{4}} 
\frac{1+\cos(\xi p\wedge k) }{((p-k)^{2}-m^{2}+i\epsilon)
(k^{2}-m^{2}+i\epsilon)}
\end{eqnarray}
Since the term without the factor $\cos(\xi p\wedge k)$ is 
identical to the conventional theory, we concentrate on the term
 with $\cos(\xi p\wedge k)$
\begin{eqnarray}
&&\frac{-g^{2}}{8}\int \frac{d^{4}k}{(2\pi)^{4}}
[e^{i\xi p\wedge k}+ e^{-i\xi p\wedge k}]
\int_{0}^{\infty}\alpha d\alpha\int_{0}^{1}dx
e^{i\alpha[k^{2}+x(1-x)p^{2}-m^{2}+i\epsilon]}
\nonumber\\
&&=\frac{i g^{2}}{4(4\pi)^{2}}
\int_{0}^{\infty}\frac{d\alpha}{\alpha}\int_{0}^{1}dx
e^{i\alpha[x(1-x)p^{2}-m^{2}+i\epsilon]-i\frac{\xi^{2}
\tilde{p}^{2}}{4\alpha}}
\end{eqnarray}
where
\begin{equation}
\tilde{p}^{\mu}=\theta^{\mu\nu}p_{\nu}.
\end{equation}
See also (3.18).
For space-time noncommutative theory with $\theta^{01}\neq 0$, 
for example,
\begin{equation}
\tilde{p}^{2}\sim p_{1}^{2}-p_{0}^{2}
\end{equation}
and for space-space noncommutative theory with 
$\theta^{23}\neq 0$, for example,
\begin{equation}
\tilde{p}^{2}\sim -p_{2}^{2}-p_{3}^{2}.
\end{equation}
We thus obtain by using the result in (3.20)
\begin{eqnarray}
&&g^{2}i\Sigma(p, \xi)_{non-planar}\nonumber\\
&&=\frac{\pi g^{2}}{4(4\pi)^{2}} H^{(2)}_{0}(-i
\sqrt{(-\xi^{2}\tilde{p}^{2})[-x(1-x)p^{2}+m^{2}-i\epsilon)})
\end{eqnarray}
for $\xi^{2}\tilde{p}^{2}<0$, namely, for space-space 
noncommutative theory. For space-time noncommutative theory, for
which $\xi^{2}\tilde{p}^{2}$ can be positive as well as 
negative, one defines the amplitude by an analytic continuation. 

As for the consistency of the naive quantization (4.7), it is 
important to analyze the self-energy correction in 
\begin{equation}
\langle T^{\star}\hat{\phi}(x)\hat{\phi}(y)\rangle_{ren}
=\int\frac{d^{4}p}{(2\pi)^{4}}
e^{-ip(x-y)}\frac{i}{p^{2}+g^{2}\Sigma(p, \xi)-m_{r}^{2}}. 
\end{equation}
If $\Sigma(p, \xi)$ contains a non-local exponential factor when
one incorporates higher order quantum corrections,
the naive quantization is not justified in the framework of the 
BJL prescription as we explained for 
the simple theory non-local in time in the previous section.
In the BJL analysis, we need to cut-off the loop momenta as 
in (3.27). 

By using the asymptotic expansion in (3.21), we have 
$\Sigma(p, \xi)$ which assumes real values and exponentially 
decreases 
\begin{eqnarray}
\Sigma(p, \xi)\sim \int_{0}^{1}dx 
\sqrt{\frac{2}{\pi z}}e^{-z}
\end{eqnarray}
with
\begin{equation}
z=\sqrt{(-\xi^{2}\tilde{p}^{2})[-x(1-x)p^{2}+m^{2}-i\epsilon)}
\end{equation}
for $p_{0}\rightarrow i\infty$ and $\xi^{2}\tilde{p}^{2}<0$, 
namely, for space-space noncommutative theory. On the other 
hand, we have a damping oscillatory behavior  
\begin{eqnarray}
\Sigma(p, \xi) \sim i \int_{0}^{1}dx
\sqrt{\frac{2}{\pi z}}e^{(- iz+i\frac{\pi}{4})}
\end{eqnarray}
with
\begin{equation}
z=\sqrt{(\xi^{2}\tilde{p}^{2})[-x(1-x)p^{2}+m^{2}-i\epsilon)}
\end{equation}
for $p_{0}\rightarrow i\infty$ and $\xi^{2}\tilde{p}^{2}>0$, 
namely, for space-time noncommutative theory.

We thus conclude that the naive quantization (4.7), either in 
space-space or in space-time noncommutative theory, is not 
modified by the one-loop corrections in the framework of BJL
prescription. This is in sharp contrast to the simple theory
non-local in time analyzed in the previous section. This 
difference arises from the fact that 
\begin{equation} 
p_{\mu}\theta^{\mu\nu}p_{\nu}=0
\end{equation}
and thus the two-point function, which depends on the single
momentum $p_{\mu}$, does not contain an extra
exponential factor in the present space-time noncommutative
theory. We thus expect that our result based on the one-loop 
diagram is valid for higher loop diagrams with a suitable cut-off
of loop-momenta. 
Although our analysis does not justify the naive 
quantization to the non-perturbative accuray, it provides a 
basis of the formal perturbative expansion in the present 
model\cite{filk}.

As for the perturbative unitarity, we observe that  
\begin{eqnarray}
S(t_{+}, t_{-})=e^{i\hat{H}_{0}t_{+}}e^{-i\hat{H}(t_{+}-t_{-})}
e^{-i\hat{H}_{0}t_{-}}
\end{eqnarray}
is unitary in the present case also
\begin{equation}
S(t_{+}, t_{-})^{\dagger}S(t_{+}, t_{-})
=S(t_{+}, t_{-})S(t_{+}, t_{-})^{\dagger}=1
\end{equation}
where the total Hamiltonian $\hat{H}=\int d^{3}x{\cal H}$ is 
defined by 
\begin{eqnarray}
{\cal H}&=&\frac{1}{2}\Pi^{2}(0,\vec{x})
+\frac{1}{2}\vec{\nabla}\phi(0,\vec{x})\vec{\nabla}
\phi(0,\vec{x})+\frac{1}{2}m^{2}\phi^{2}(0,\vec{x})\nonumber\\
&+&\frac{g}{2\cdot 3!}[\phi(0,\vec{x})\star\phi(0,\vec{x})\star
\phi(0,\vec{x}) + h.c. ]
\end{eqnarray}
with the naive canonical momentum $\Pi(x)
=\frac{\partial}{\partial x^{0}}\phi(x)$ conjugate to the 
variable $\phi(x)$. See \cite{gomis} for a different approach 
to the Hamiltonian formulation of space-time noncommutative 
theory. 
The operator defined by a formal perturbative expansion of (4.20)
\begin{eqnarray}
\hat{S}&=&\sum_{n=0}^{\infty}\frac{(-i)^{n}}{n!}
\int_{-\infty}^{+\infty}dt_{1}....\int_{-\infty}^{+\infty}dt_{n}
T_{\star}(\hat{H}_{I}(t_{1})....\hat{H}_{I}(t_{n}))\nonumber\\
&=&\sum_{n=0}^{\infty}(-i)^{n}
\int_{-\infty}^{+\infty}dt_{1}....\int_{-\infty}^{+\infty}dt_{n}
\theta(t_{1}-t_{2})...\theta(t_{n-1}-t_{n})
(\hat{H}_{I}(t_{1})....\hat{H}_{I}(t_{n}))\nonumber\\
\end{eqnarray}
with
\begin{eqnarray}
\hat{H}_{I}(t)&\equiv&e^{i\hat{H}_{0}t}\int d^{3}x
\frac{g}{2\cdot 3!}[\phi(0,\vec{x})\star\phi(0,\vec{x})\star
\phi(0,\vec{x}) + h.c. ]
e^{-i\hat{H}_{0}t}\nonumber\\
&=&\int d^{3}x
\frac{g}{2\cdot 3!}[\phi(t,\vec{x})\star\phi(t,\vec{x})\star
\phi(t,\vec{x}) + h.c. ]
\end{eqnarray}
and 
\begin{equation}
H_{0}=\int d^{3}x[\frac{1}{2}\Pi^{2}(0,\vec{x})
+\frac{1}{2}\vec{\nabla}\phi(0,\vec{x})\vec{\nabla}
\phi(0,\vec{x})    
+\frac{1}{2}m^{2}\phi^{2}(0,\vec{x})]
\end{equation}
defines a unitary S-matrix
\begin{equation}
\hat{S}\hat{S}^{\dagger}=\hat{S}^{\dagger}\hat{S}=1.
\end{equation}
Note that the time-ordering in (4.23) is defined with repsect to
 the time $t$ of $\hat{H}_{I}(t)$. 
Because of the Moyal product, the interaction Hamiltonian
$\hat{H}_{I}(t)$ is not local in the time variable. We thus 
encounter the same complications as in the non-local theory 
we analyzed in the previous section. The unitary S-matrix
thus generally spoils the perturbative positive energy 
condition. 

On the other hand, the conventional S-matrix, which corresponds 
to the one given by the path integral, 
\begin{eqnarray}
\hat{S}=\sum_{n=0}^{\infty}\frac{(-i)^{n}}{n!}
\int_{-\infty}^{+\infty}dt_{1}....\int_{-\infty}^{+\infty}dt_{n}
T^{\star}(\hat{H}_{I}(t_{1})....\hat{H}_{I}(t_{n}))
\end{eqnarray}
is based on the time ordering of the time variable appearing in 
each field variable $\phi(t,\vec{x})$ and, for example, the 
second order term given by the path integral has the property
\begin{eqnarray} 
&&\frac{1}{2}\int_{-\infty}^{\infty} 
\int_{-\infty}^{\infty} 
dt_{1}dt_{2}T^{\star}\hat{H}_{I}(t_{1})\hat{H}_{I}(t_{2})
\nonumber\\
&\neq&\int_{-\infty}^{\infty} 
\int_{-\infty}^{\infty} 
dt_{1}dt_{2}\theta(t_{1}-t_{2})
\hat{H}_{I}(t_{1})\hat{H}_{I}(t_{2})
\end{eqnarray}
for the space-time noncommutative theory and thus it is not 
unitary, though the positive energy condition in the sense that 
the positive energy always flows in the positive time direction 
is satisfied. We emphasize that (4.27) is defined directly by
the path integral without recourse to the expression such as 
(4.20).

To be more explicit, we have 
\begin{eqnarray}
&&\frac{1}{2}[\phi(x)\star\phi(x)\star\phi(x)+ h.c.]\nonumber\\
&&=\cos(\frac{\xi}{2}(\partial_{x_{1}}\wedge(\partial_{x_{2}}+
\partial_{x_{3}})+\partial_{x_{2}}\wedge\partial_{x_{3}}))
\phi(x_{1})\phi(x_{2})\phi(x_{3})|_{x_{1}=x_{2}=x_{3}=x}
\nonumber\\
&&=\sum_{p_{1},p_{2},p_{3}}\cos(\frac{\xi}{2}(p_{1}\wedge(
p_{2}+p_{3}) + p_{2}\wedge p_{3}))
e^{ip_{1}x}e^{ip_{2}x}e^{ip_{3}x}
\phi(p_{1})\phi(p_{2})\phi(p_{3})
\nonumber\\
&&=\sum_{p_{1},p_{2},p_{3}}\frac{1}{2}
[e^{ip_{1}(x+\frac{\xi}{2}\wedge(
p_{2}+p_{3}))}e^{ip_{2}(x+\frac{\xi}{2}\wedge p_{3})}
e^{ip_{3}x} + e^{ip_{1}(x-\frac{\xi}{2}\wedge(
p_{2}+p_{3}))}e^{ip_{2}(x-\frac{\xi}{2}\wedge p_{3})}
e^{ip_{3}x}]\nonumber\\
&&\times
\phi(p_{1})\phi(p_{2})\phi(p_{3})
\end{eqnarray}
Although the way of writing the last line of the above equation
is not unique,
it shows that the non-local parameter in the present context is 
momentum dependent and non-locality becomes more significant 
for the larger momenta of the neighboring fields. 

By using this interaction vertex (4.29), the evaluation of the 
one-loop self-energy diagram on the basis of the conventional 
$T$ or $T^{\star}$ product starts with 
\begin{eqnarray}
&&-\frac{1}{2}\frac{g^{2}}{(3!)^{2}}\int_{-\infty}^{\infty} 
d^{4}x d^{4}y T^{\star}\frac{1}{2}
[\phi(x)\star\phi(x)\star\phi(x)+h.c.]\nonumber\\
&&\ \ \ \ \ \ \ \ \ \ 
\times \frac{1}{2}[
\phi(y)\star\phi(y)\star\phi(y)+h.c.]
\end{eqnarray}
and this gives rise to 
\begin{eqnarray}
\frac{g^{2}}{2}\int \frac{d^{4}l}{(2\pi)^{4}} 
\frac{\cos^{2}(\frac{\xi}{2}p\wedge l)}
{((p-l)^{2}-m^{2}+i\epsilon)(l^{2}-m^{2}+i\epsilon)}
\end{eqnarray}
after integrating over the coordinates of the two vertex points,
 as we have discussed in (4.8).
The Feynman's $m^{2}-i\epsilon$ prescription ensures the 
conventional time ordering of each field variable $\phi(x)$ by 
taking into account the momentum 
dependent non-local effects; the positive energy always flows in
 the forward time direction by incorporating  the momentum 
dependent non-local effects. It is known that the present 
expression of the one-loop two point function (4.31) does not 
satisfy
the unitarity relation\cite{gomis-mehen, alvarez-gaume, chu}, as 
is witnessed by the non-vanishing imaginary part of 
$\Sigma(p, \xi)$ in (4.17) for the Euclidean momentum given by 
$p_{0}\rightarrow i\infty$.
The conventional Wick rotation to Euclidean theory is 
well-defined because of the $m^{2}-i\epsilon$ prescription, if 
one defines a suitable rotation of the wedge product $p\wedge l$.

On the other hand, the non-local prescription starts with
\begin{eqnarray}
&&-\frac{g^{2}}{(3!)^{2}}\int_{-\infty}^{\infty} 
d^{4}x d^{4}y\theta(x^{0}-y^{0})
\frac{1}{2}[\phi(x)\star\phi(x)\star\phi(x)+h.c.]\nonumber\\
&&\ \ \ \ \ \ \ \ \ 
\times \frac{1}{2}[\phi(y)\star\phi(y)\star\phi(y)+h.c.]
\end{eqnarray}
and the representation
\begin{equation}
\theta(x^{0}-y^{0})=-\int_{-\infty}^{\infty}
\frac{d\omega}{2\pi i}
\frac{e^{-i\omega(x^{0}-y^{0})}}{\omega+i\epsilon}.
\end{equation}
We thus obtain after extracting the overall 4-momentum conserving
$\delta$-function
\begin{eqnarray}
\frac{g^{2}}{2}\int \frac{d^{4}l}{(2\pi)^{4}}
\int_{-\infty}^{\infty}\frac{d\omega}{2\pi i(\omega+i\epsilon)} 
\cos^{2}(\frac{\xi}{2}p\wedge l)\tilde{\Delta}_{+}(l)
\tilde{\Delta}_{+}(p-l-\omega)
\end{eqnarray}
where 
\begin{equation}
\Delta_{+}(x-y)=\langle 0|\phi(x)\phi(y)|0\rangle
=\int\frac{d^{4}k}{(2\pi)^{4}}e^{-ik(x-y)}\tilde{\Delta}_{+}(k)
\end{equation}
and $\tilde{\Delta}_{+}(k)=2\pi\delta(k^{2}-m^{2})\theta(k^{0})$
; the variable $\omega$ in $p-l-\omega$ stands for 
$(\omega,0,0,0)$. It is known that this expression (4.34)
satisfies the unitarity relation\cite{bahns, rim-yee} though the
 energy-momentum 
conservation, which is a result of the translational invariance 
of the starting action, is not manifest in the present notation.
The time ordering in the present case (4.32) is specified by 
$\theta(x^{0}-y^{0})$ in 
front of the Moyal products, and thus the time ordering of each 
field variable $\phi(x)$ induced by the space-time 
noncommutative product is ignored. Also, a smooth Wick 
rotation to Euclidean theory is not obvious.

\section{Discussion}

We illustrated that  the path integral on the basis of 
Schwinger's action principle  has a wide range of applications.
The time ordering of field operators is rigidly specified to be 
the conventional one in the path integral. In this sense the 
path integral has little flexibility as to  modified time 
ordering
operations. In some examples such as higher derivative 
theory, we have shown that  the canonical commutation relations 
are readily recovered from the correlation functions defined by 
the path integral.    

We analyzed some of the basic aspects of quantized theory
which is non-local in the time variable on the basis of the path
integral quantization. In general, the naive quantization in the 
sense of the interaction picture is not justified, but we showed 
that the space-time noncommutative theory is stable under higher
 order quantum corrections in a perturbative sense in sharp 
contrast to a naive theory non-local in time. Although we 
analyzed this issue for a simple scalar theory, we expect that 
the conclusion is valid for a more general class of field 
theories and thus this provides a basis for a perturbative 
analysis of space-time noncommutative theory. 

In view  of various time-ordering operations 
available in the operator formulation, we analyzed the recent 
proposal of the modified time ordering prescription\cite{bahns,
rim-yee}, which generally defines 
a unitary S-matrix for a  theory non-local in time variable.
This freedom of the modified time ordering is not available for 
the path integral, and thus 
specific to the operator formulation. We showed that the unitary
S-matrix has certain advantages but at the time it has several
 disadvantages, and the perturbative positive energy condition 
and a smooth Wick rotation to Euclidean theory, which are 
ensured by the Feynman's $m^{2}-i\epsilon$ prescription, are 
spoiled. 
       
Since a  quantization scheme of space-time noncommutative theory
 satisfactory in every respect is not known at this moment, 
our conclusion is that the path integral 
quantization scheme with Feynman's $m^{2}-i\epsilon$ 
prescription is attractive, which is simple in principle 
and allows a smooth definition of Euclidean theory indispensable
 for some non-perturbative analyses. The path integral 
formulation displays the difficulty of the space-time 
noncommutative theory as an absence of the unitary S-matrix.
 
As for the compelling motivation for studying the 
noncommutative space and time in fundamental physics, one may 
count the recent developments related to string theory such as 
in \cite{connes-douglas-schwarz, seiberg-witten, aoki}, for 
example, but the analysis of such concrete examples is beyond 
the scope of the present paper.

\end{document}